# A Gd@C$_{82}$-based single molecular electret device

*Kangkang Zhang[1†], Cong Wang[2†], Minhao Zhang[1†], Zhanbin Bai[1†], Fangfang Xie[3], Yuanzhi Tan[3], Yilv Guo[4], Kuo-Juei Hu[1], Lu Cao[1], Shuai Zhang[1], Xuecou Tu[5], Lin Kang[5], Jian Chen[5], Peiheng Wu[5], Xuefeng Wang[5], Jinlan Wang[4], Junming Liu[1], Baigeng Wang[1], Guanghou Wang[1], Suyuan Xie[3*], Wei Ji[2*], Su-Fei Shi[6,7*], M. A. Reed[8,9*], Fengqi Song[1,9*]*

[1] *National Laboratory of Solid State Microstructures, Collaborative Innovation Center of Advanced Microstructures, and School of Physics, Nanjing University, Nanjing 210093, China*

[2] *Beijing Key Laboratory of Optoelectronic Functional Materials & Micro-Nano Devices, Department of Physics, Renmin University of China, Beijing 100872, China*

[3] *State Key Laboratory of Physical Chemistry of Solid Surfaces and Department of Chemistry, College of Chemistry and Chemical Engineering, Xiamen University, Xiamen 361005, China*

[4] *School of Physics, Southeast University, Nanjing 211189, China*

[5] *School of Electronic Science and Engineering and Collaborative Innovation Center of Advanced Microstructures, Nanjing University, Nanjing 210023, China*

[6] *Department of Chemical and Biological Engineering, Rensselaer Polytechnic Institute, Troy, New York 12180, USA*

[7] *Department of Electrical, Computer and Systems Engineering, Rensselaer Polytechnic Institute, Troy, New York 12180, USA*

*[8]Departments of Applied Physics and Electrical Engineering, Yale University, New Haven, CT 06520, USA*

*[9] Atomic Manufacture Institute (AMI), 211805 Nanjing, China*

*[†]These authors contributed equally to this work.*

*Correspondence and requests for materials should be addressed to F. S. (email: songfengqi@nju.edu.cn); S. X. (email: syxie@xmu.edu.cn); W. J. (email: wji@ruc.edu.cn); S.-F. S. (email: shis2@rpi.edu); M. R. (email: mark.reed@yale.edu).

**Abstract**

Single molecular electrets exhibiting single-molecule electric polarization switching have been long desired as a platform for extremely small non-volatile storage devices, although it is controversial because of the poor stability of single molecular electric dipoles. Here we study the single molecular device of Gd@$C_{82}$, where the encapsulated Gd atom forms a charge center, and we have observed a gate-controlled switching behavior between two sets of single-electron-transport stability diagrams. The switching is operated in a hysteresis loop with a coercive gate field of around 0.5 V/nm. Theoretical calculations have assigned the two conductance diagrams to corresponding energy levels of two states that the Gd atom is trapped at two different sites of the $C_{82}$ cage, which possess two different permanent electrical dipole orientations. The two dipole states are stabilized by the anisotropic energy and separated by a transition energy barrier of 70 meV. Such switching is then accessed to the electric field driven reorientation of individual dipole while overcoming the barriers by the coercive gate field, and demonstrates the creation of a single molecular electret.

**Introduction**

Moore's law pushes the miniaturization of modern electrical devices, and such a ride has been approaching the single atomic or molecular scale [1], which has ignited the research of single molecular devices (SMDs). Many efforts have been made to fabricate SMDs and design functional devices such as switches [2-7], diodes [8-11], transistors [12-16] and sensors [17-19], which are believed to be the building blocks of future molecular computers [20]. Among all molecules which provide versatility in the design of SMDs, fullerenes and their derivatives stand out [12,21] due to their unique cage-like structures. $C_{60}$-based SMDs have shown numerous interesting physics and striking properties, such as the Kondo effect [22], quantum phase transition [23], superconductivity [24], negative differential resistance [25] and other effects [26]. As a second strategy, foreign atoms can be introduced into fullerene cages to cast new physics, in which charge centers and magnetic centers have been studied, and superconductivity can even be strengthened. A recent study even reports the protection of quantum information of N atoms by the cage with a coherence time of up to 0.23 ms [27]. As a further dimension to control the physical properties, the modulation of the cages ($C_{70}$, $C_{82}$, etc) have also been explored. For examples, $C_{82}$ has a deformed cage with lower symmetry and RE@$C_{82}$ (RE = La, Gd, Tb, Dy, Ho, Er) have been prepared with the aim of extremely small and stable storage units [28-31], where the SMD of Ce@$C_{82}$ [32] has ever accessed an additional state. Such continuous molecular optimization efforts finally lead to non-superimposable charge center and structural center [33,34], which may form stabilized electric dipoles in the molecule. This can enable the long desired single molecular electret (SME), which we show in this work.

We report the gate-controlled switching between two sets of characteristic single-electron stability diagrams in the electrical transport of a Gd@$C_{82}$-based SMD. It is fabricated by a feedback-controlled electromigration break junction (FCEBJ) method [12] and can be operated in a hysteresis-like loop with a coercive gate field of around 0.5 V/nm. The supporting theoretical calculations interpret the switching as the electrical-field-driven reorientation of single molecular dipoles anddemonstrates for the first time single molecule electret switching devices.

**Single electron transport in Gd@$C_{82}$ transistors**

As schematically shown in **Figure 1(a)**, the single Gd@$C_{82}$ molecular transistor is prepared by setting a Gd@$C_{82}$ molecule into a pair of Au electrodes that have a nanometer-scale gap, which is fabricated according to the details provided in the methods section. Fig. 1(b) shows a set of *I-V* curves in a typical FCEBJ procedure[12] and the device's morphology is shown in the inset. Great attention has been paid during the thermal oxide growth in order to achieve a satisfactory gating efficiency and local gating field, which are critical in this Gd@$C_{82}$ study with a sizeable density of states near the Fermi level. We etch the previously thick $SiO_2$ using dilute HF, then regrow a new layer of $SiO_2$ about 30nm by dry oxidation method, repeated twice.

As shown in Fig. 1(c), a back gate of 7-10 V can modulate the *I-V* curve. A nonlinear dependence between the current and the voltage can be seen, and there is a current blockade area at low voltage. The gate voltage can decrease or increase the blockade voltage range, indicating change in the electrochemical potential. A plot of the differential conductance d$I$/d$V$ as a function of the bias voltage $V_{ds}$ and the gate voltage $V_g$ is shown in Fig. 1(d). The differential conductance is derived by numerical differentiation of the current with respect to the bias voltage. The blue regions are the Coulomb blockade regions, and the red regions show differential conductance peaks. For clarity, two black dashed lines are shown in Fig. 1(d) to indicate the SMD Coulomb edges. Changing the gate voltage will tune the chemical potential of the molecule. When the electrochemical potential is aligned with the Fermi energy of the source and drain electrodes, the conductance gap will decrease to zero and reach a degeneracy point (where the two black dashed lines cross), indicating the successful preparation of a single electron with Gd@$C_{82}$ SMD. With a larger gate voltage scan, multiple peaks are observed as shown in **Figure S1**, indicating that our SMD could access a series of redox states and degeneracy points. Such multiple degeneracy points have been previously related to the molecular energy levels [16].

**Gate-controlled switching between two electronic states and its hysteresis loop**

An interesting switching between two molecular states is observed. We carefully measure the current as a function of the gate voltage when the source-drain bias is fixed at 2 mV as shown in **Figure 2(a)**. The gate voltage is swept backwards and forwards over a large range. In contrast to the literature, the source-drain current shows two sets of coulomb oscillation patterns. One may see the red and green curves are similar, defined as **State 1** and the blue curve is defined as **State 2**. The two sets of coulomb oscillation patterns can be reversibly switched, in contrast to the previous result in Ce@$C_{82}$[32], when after application of a large bias voltage,the Coulomb stability diagram is irreversibly changed. Furthermore, the oscillation peaks can be related to the molecular orbitals (as shown later), hence we speculate that the onset of two sets of oscillation peaks means that Gd@$C_{82}$ may have two metastable states. Our observation suggests that the structure of the molecule can be controlled and switched by the external electric field.

Systematic two-dimensional plots of the differential conductance are measured over a large range of gate voltage. The detailed Coulomb stability diagrams of the two states are shown in Fig. 2(d) and 2(e) respectively, which show a distinct difference. Firstly, the Coulomb degeneracy points are different, corresponding to the Coulomb oscillations at different gate voltages. Secondly, the slopes of the Coulomb edges, i.e., the white line in the plots (black dashed lines are shown in Fig. 2(d) and 2(e) for clarify), are also different. We investigate these differences between the two states quantitatively by using a capacitance model to calculate the SMD's detail parameters[35]. In Fig. 2(d), the $C_G$ : $C_S$ : $C_D$ ratio is 1:15.05:133.23, and the gate efficiency factor α changes to 0.0067 (where e$\alpha V_g$ is the change in the electrochemical potential caused by the gate electrode). In Fig. 2(e), the $C_G$: $C_S$: $C_D$ ratio for the three capacitors is 1:15.34:86.97 and the gate efficiency factor α is 0.0097. The drastic difference in both gate efficiency and gate capacitances confirms that the SMD undergoes a significant change of its electronic states. The difference between these two stability diagrams has it's origin in the order in which the gate voltage is applied.

Starting from the state of Fig. 2(d) ("State 1"), the gate voltage is first decreased to -13 V, and then changed up to 15 V, leading to Fig. 2(e) ("State 2"). To emphasize the differences between the two states, , two cut lines of Fig. 2(d) and 2(e) (red and black curves, respectively) are shown in Fig. 2(f) when the bias voltage is zero.. Two subsection plots shown in **Figure S2**, illustrating the differences between these two stability diagrams. These two sets of stability diagrams are highly repeatable, reproducible, and stable for more than one month, indicating that the gate voltage changes the internal structure of the molecule. From the two Coulomb stability diagrams, we can gain further information about these two metastable states of Gd@$C_{82}$ with the aid of theory calculations as shown below.

Repeatable switching between the two states results in a ferroelectricity-like hysteresis loop with a coercive gate field of around 0.5 V/nm. Analyzing the data shown in Fig. 2(a), we summarize the switching of the SMD device as shown in Fig. 2(b), where the thin blue-arrowed line indicates the gate application sequence. We respectively label the diagram of Fig. 2(d) as the dark arrow (State 1) and the diagram of Fig. 2(e) as the orange arrow (State 2). The starting state is initially set to State 1 when the gate voltage is 0 V. The device maintains State 1 even if we decrease the gate voltage to -10 V. We then change the gate voltage direction. When the gate voltage is increased positively to over 10.5 V, the device was found to switch to State 2. The state can be set back to State 1 when a large enough negative gate voltage is applied. Such switching/reversal can be reliably reproduced as seen in Fig. 2(b), which can be understood by the state switching of the molecule driven by the out-of-plane electric field provided by the back-gate voltage. Fig. 2(b) and (c) essentially depict a hysteresis loop in typical ferroelectricity operations at single molecular level, essentially the evidence of SME.

The reliable switching of the two states leads to a series of programmable operations. In fact, any gate voltage point in Fig. 2(a) can be employed to simulate a two-resistance-state operation and SMD storage, as long as each point has a different current. We select one point in Fig. 2(a) as an example. In one point of the state, the current is very low because the SMD was in the blockade state, while in another molecular state the current is much higher when the blockade is removed. This operation is shown in Fig. 3(a). The switch gate voltage is ±11 V and the test gate voltage is 7.472 V to read the molecular state, as shown in the bottom panel of Fig. 3(a). The corresponding current as a function of time is shown in the top panel of Fig. 3(a), with a bias voltage of 2 mV. The 'test' current takes on two values, marked 'low' and 'high' at the top Fig. 3(a). Different currents (~62% difference) indicate different molecular states, which can be used in a typical 2-state electret storage device. The time step could be changed to 10 or 1 s, as shown in Fig. 3(b), in which the red line is the bias current. It can clearly be seen that there are two obvious different current at the test gate voltage no matter what time step is used, and this observation can be repeated many times. This confirms that the single Gd@$C_{82}$ molecular device has two metastable states and can be reliably set and reset by a gate voltage, which we relate to a controlled switching of SMEs as shown below.

**Theoretical modeling revealing the SME physics**

Density functional theory calculations were performed to unveil the corresponding atomic configurations of the two switchable states. Details of our calculations can be found in the method and supplementary section. While totally seven configurations were considered, as shown in supplementary **Figure S4**, two of them, denoted configurations I and IV respectively, are significantly more stable, by over 60 meV than the other five ones. **Figure 4(a)** shows the geometric details of configurations I and IV where the Gd atom sits over two neighboring $C_6$ hexagonal rings as marked with red and blue colored hexagons, respectively. In a neutral state, configuration I is roughly 22 meV more stable than configuration IV. This energy difference reduces under either electron or hole doping and turns to negative values at a hole doping level of over 1.5 $h/C_{82}$ (Fig. 4(b)), which indicates switchable relative stability of configurations I and IV.

The energy levels for these two states could be extracted using a capacitance model from the data shown in Figs. 2(d) and 2(e). We have previously calculated two gate efficiency factors $\alpha$, where $e\alpha V_g$ is the change in the electrochemical potential caused by the gate electrode. The energy difference between two adjacent states plus the Coulomb energy was defined as a pair of adjacent degeneracy points[16] using equation $\alpha e(V_G(\text{n}) - V_G(\text{n}+1)) = \varepsilon^{n+1} - \varepsilon^n + e^2/(C_s + C_d + C_g)$. The molecular energy levels are extracted from the degeneracy points of Fig. 2(e) and 2(d) and are marked in Figs. 4(c) and 4(d), respectively. Figs. 4(e) and 4(f) show the eigenvalues of occupied states of configurations I and IV (with a hole doping level of 2 $h/C_{82}$), respectively. The derived molecular levels and the theoretical eigenvalues are reasonably well consistent unless some minor features, which are probably due to contact induced defects. The comparison of molecular levels and eigenstate values strengthens the assignment of configuration I and IV being the two observed states, respectively.

Fig. 4(g) plots the calculated transition pathways from configurations I to IV at different hole doping levels. It shows the energy barrier reduces from 78 meV of the neutral state to 45 meV at the 2.0 $h/C_{82}$ doping level, accompanying with the enhanced stability of configuration IV. Given the derived gate efficiencies and the transition voltages, we could infer a potential energy of 87 meV was used by the measured device in transferring configuration I to IV while the opposite transition needs an energy of 70 meV, as illustrated in Fig. 4(a). These values, i.e. 70~87 meV, are highly consistent with the theoretical transition barriers of 45-78 meV, which well explains the onset of the coercive gate field in Fig. 2(b).

Either applied external electric field or its induced charge doping helps the Gd atom to surmount this barrier. In both configurations, the Gd atom is prone to adsorb over $C_6$ rings rather than at the center of the cage, resulting in stabilized electric dipole moments, i.e. SME. The calculated dipole moments of configurations I and IV are respectively illustrated in Fig. 4(a), in which red and light blue arrows represent the direction of the dipole moments in the two configurations, respectively. Details of calculated dipole moments are available in Table S3. Since a lateral source-drain voltage is applied when forming the molecular junction, we align the moment of configuration I of 0.50 $e\cdot$Å to the x direction. A vertical moment of 0.37 $e\cdot$Å, parallel to the z direction, emerges after the transition to configuration IV. The switching between the two states can be simplified as an electric-field driven flipping of SME of Gd@$C_{82}$, which requires an electric field of ~120 mV/Å to surmount the predicted energy barrier of ~45 meV. The estimated coercive field of ~ 120 mV/Å is within the range of the experimentally applied field to observe the ferroelectricity (electret)-featured hysteresis loop where the transition occurs when 10~15 V voltages act on a 10~30 nm-thick gate layer (33 ~ 150 mV/Å).

The physics of such SME is different from that of ferroelectricity in solids, where the system is stabilized by the exchange coupling between huge numbers of dipoles in a large domain. It is more alike to the single molecule magnet, in which spin polarization is stabilized by anisotropy energy of individual molecules. Such SME free of any inter-dipole coupling has been anticipated by some recent theories[36] and experimentally tested[20], where the sizable inter-dipole coupling is still seen in a long-range ordered crystal and casts a question on this topic. The present experiment demonstrates the SME at the single molecular level and thus excludes any inter-dipole coupling, which forms convincing evidence of a SME. The unique versatility of the metastable structure of RE@$C_{82}$ enables the observation of the SME in SMDs. This may pave the way towards the extremely small storage devices in future electronics.

## Method

### Molecules preparation

Soot containing Gd@$C_{82}$ is produced using the direct current arc discharge method with metal/graphite composite rods in a helium atmosphere. The soot is extracted with carbon disulfide and the residue is extracted further with pyridine under reflux. Gd@$C_{82}$ molecules are extracted from the cleaned soot using a two-stage high-performance liquid chromatography method. Stage 1 is a preliminary high-performance liquid chromatography process using a Buckyprep column. This yields a fraction containing Gd@$C_{82}$. The second stage involves isolation of Gd@$C_{82}$ from empty fullerenes using a Buckyprep-m column.

### Devices preparation

The ~50 nm wide gold nanowires are deposited on top of a silicon gate electrode that has previously been oxidized in an oxygen atmosphere to give a $SiO_2$ layer (~30 nm thick) to act as a gate dielectric. The electrodes and $SiO_2$ insulator are patterned using an ordinary photolithography procedure, but the nanowire is defined by electron-beam lithography, as shown in Fig. 1(b) inset. When the Au nanowires are prepared, to make sure the chip is clean, the electrodes are cleaned further using oxygen plasma. Then the Gd@$C_{82}$ contained device is cooled to 1.6 K, and a nanogap is produced using the FCEBJ method [12]. Briefly to say, the current is monitored while the applied voltage between the source and drain is increased. The voltage is decreased to 10 mV as soon as the current drops by 1%. The cycle is repeated until the resistance is >1 MΩ when the voltage is 20 mV.

### DFT calculations

Our density functional theory calculations are performed using the generalized gradient approximation and the projector augmented wave method [37,38] as implemented in the Vienna *ab-initio* simulation package (VASP) [39]. The kinetic energy cut-off is set to 400 eV. An 18 × 18 × 18 Å$^3$ supercell is used to model the isolated Gd@C$_{82}$ molecule, which ensures a separation of at least 7.5 Å among image molecules. This separation, as previously demonstrated, is sufficient to reduce appreciably image interactions. The Gamma point is used for sampling the first Brillion zone in all calculations. All atoms are allowed to relax until the residual force on each atom is less than 0.01 eV/Å. Dispersion interactions are considered at the van der Waals density functional (vdW-DF) level [40-42], with the optB86b functional for the exchange potential (optB86b-vdW), which is proved to be accurate in describing the structural properties of layered materials and is adopted for structure-related calculations[43-47]. Transition pathways and energy barriers are revealed by the climbing image nudged elastic band (CI-NEB) method [48,49], which locates the exact saddle point of a reaction pathway. Electric dipole moments of the C$_{82}$ molecule system are calculated on the basis of the classical definition:

$$P = \frac{1}{V}\left(-e\sum_j Z_j \boldsymbol{u}_j\right) + \int \boldsymbol{r}\rho(r)d\boldsymbol{r}$$

Here $e$ is the electron charge, $V$ is the cell volume, $Z_j$ and $\boldsymbol{u}_j$ are the atomic number and the position of the $j_{th}$ atom, $\rho(r)$ is the electronic charge density. Dipole correction is considered in all calculations to correct the error introduced by the periodic boundary condition and balance the vacuum level differences on the different sides of the polarized molecules[50,51]. All electronic structures are calculated using the

Perdew-Burke-Ernzerhof (PBE) functional [52] with inclusion of spin-orbit coupling (SOC) on the basis of the optB86b-vdW revealed structures. An on-site Coulomb interaction energy of 6 eV is added to the $f$ orbitals of the Gd atom according to the values reported in the literature[53-55].

**Analysis of structures**

We first consider seven adsorption sites of a single Gd atom adsorption on the inner wall of a $C_{82}$ molecule (Fig. S3 and Table S1). Among them, the $C_6$-Hol site where a Gd atom sitting over a $C_6$ hexagonal ring (Fig. S4) shows superior stability that its energy is over 1 eV more stable than the other two 6C-ring sites and the adsorbed configurations of other sites transform into the mentioned $C_6$-Hol site (Table S1). There are seven inequivalent $C_6$-rings of a $C_{82}$ molecule as shown in Fig. S4, we next double-check relative stability of these seven configurations. The relative total energy and the height of the adsorbed Gd atom at every site are listed in Table S2. We denote the $C_6$-Hol configuration we previously obtained as configuration I, while others are represented as configurations II toVII, as shown in Fig. S4. Configuration I is, again, the most stable one among these seven configurations, while configuration IV is the secondarily stable one. Configuration III, the third-stable one, is 65.8 meV/$C_{82}$ less stable than configuration IV, while other configurations are at least 400 meV less favored. We notice that there are three inequivalent $C_5$ rings and also double-check the stability of being adsorbed on the other previously unconsidered $C_5$ rings. Those rings do not offer a more stable configuration (Fig. S5).

**Acknowledgments**

We gratefully acknowledge the financial support of the National Key R&D Program of China (2017YFA0303203 and 2018YFE0202700), the National Natural Science Foundation of China (91622115, 11522432, 11574217, U1732273, U1732159, 61822403, 11874203, 11904165, 11904166, 11622437, 61674171 and 11974422), the Strategic Priority Research Program of Chinese Academy of Sciences (Grant No. XDB30000000), the Fundamental Research Funds for the Central Universities, China, and the Research Funds of Renmin University of China [Grants No. 16XNLQ01 (W.J.), the Natural Science Foundation of Jiangsu Province (BK20160659), the Fundamental Research Funds for the Central Universities, and the opening Project of the Wuhan National High Magnetic Field Center. Cong Wang was supported by the Outstanding Innovative Talents Cultivation Funded Programs 2017 of Renmin University of China. Calculations were performed at the Physics Lab of High-Performance Computing of Renmin University of China and Shanghai Supercomputer Center. S.-F. Shi acknowledges support from NYSTAR through Focus Center-NY–RPI Contract C150117. We thank Prof. Hong-Ling Cai and You Song from Nanjing University for stimulating discussions.

**Additional information**

Competing interests: The authors declare no competing financial interests.

**Figure captions**

**Figure 1. Single electron transport of the Gd@C$_{82}$ SMD. a**: The SMD configuration. A single Gd@C$_{82}$ molecule bridges a nanogap in a pair of bowtie gold electrodes on top of a 30 nm thick silicon dioxide layer. **b**: A typical feedback-controlled electromigration process. The inset is a ccanning electron microscope (SEM) image of the Au nanoelectrode before it was broken. Scale bar: 100 nm. **c**: Representative $I(V)$ curves for different gate voltages $V_g$ after the electromigration. **d**: Differential conductance d$I$/d$V$ plotted against the bias voltage $V_d$ and the gate voltage $V_g$. The blue regions are the Coulomb blockade areas, and the red lines show peaks in the differential conductance. This shows the operation from N to N+1 electrons.

**Figure 2. Gate-controlled switching between the two molecular states showing a ferroelectricity-like hysteresis loop. a**: The source-drain current is plotted as a function of the gate voltage with the source-drain bias fixed at 2 mV when the gate voltage was swept backwards and forwards. The black dashed line marks several resistance points, indicating two states are accessed. State 1: the red and green curves with low currents while crossing the dashed line. State 2: other curves like the blue curve with high current while crossing the dashed line. An offset is added to each line for clarity. **b**: Schematic of the switching process. A repeatedly cycled gate voltage switches the SMD between State 1 and 2, whose conductance diagrams are also depicted detailed in panel **d** and **e** respectively. The arrowed blue line indicates the sequence that the gate voltage is applied. The molecule cannot switch between states

unless a large enough gate voltage $V_{g\ trans-}/V_{g\ trans+}$ is applied to the SMD. This indicates the onset of a coercive gate field and leads to a ferroelectricity-like hysteresis loop as shown in panel **c**. **d** and **e**: Color plot of the differential conductance d$I$/d$V$ plotted against the bias voltage $V_{ds}$ and gate voltage $V_g$, respectively, over a large voltage range for the device. From -13 V decreases to 15 V in **d**, reverse in **e**. **f**: A cut line plot of the stability diagram when the bias is 0 V. The red and black lines are extracted from **d** and **e**, respectively. Arbitrary units and an offset are employed for clarity.

**Figure 3. Simulating a two-resistance-state operation based on the SMD switching. a**: The bottom figure is the gate voltage plotted against time. The switch gate voltage is ±11 V and the test gate voltage is 7.472 V, using a time step of 60 s. The top figure is the current plotted against the same time as used in the bottom figure, with the bias voltage at 2 mV. The 'test' current switches between the two parts marked 'low' and 'high'. **b**: The switch between two molecular states can be achieved using time steps of 60, 10, and 1 s. The black and red lines are the gate voltage and bias current, respectively.

**Figure 4. Density functional theory calculations revealing the SME physics. a**: The switching is interpreted as a transition between $C_6$-Hol-I and $C_6$-Hol-IV. The left and right are the two configurations ($C_6$-Hol-I, $C_6$-Hol-IV) respectively. Two colored arrows represent the direction of the dipoles. Changing the electrochemical potential of the Gd@$C_{82}$ SMD by >70 meV transforms the molecule from $C_6$-Hol-IV to $C_6$-Hol-I, which will be maintained until a reverse shift >−87 meV occurs. This can be simplified as the gate field driven flipping of the electric dipole of single Gd@$C_{82}$ molecule, i.e. SME. **b**: Relative total energy of the two configurations as a function of electron/hole doping levels. The total energy of $C_6$-Hol-I is set as zero. **c** and **d**: Differential conductance plotted against the effective molecular orbital gating α$eV_g$ with the bias voltage $V_{ds}$ at 0 V. Extracted from the experimental data of Fig 3(e) and (d) respectively. **e** and **f**: Calculated Eigenvalues of occupied states of the neutral $C_6$-Hol-I and hole-doped (2$h$) $C_6$-Hol-IV, respectively. The energy scale is relative to the vacuum level. The energy of the Fermi levels are -5.07 eV and -7.06 eV in *e* and *f*, respectively. These agree well with the experimental data in panels c and d. **g:** Calculated transition paths and energy barrier between $C_6$-Hol-I and $C_6$-Hol-IV upon hole doping.

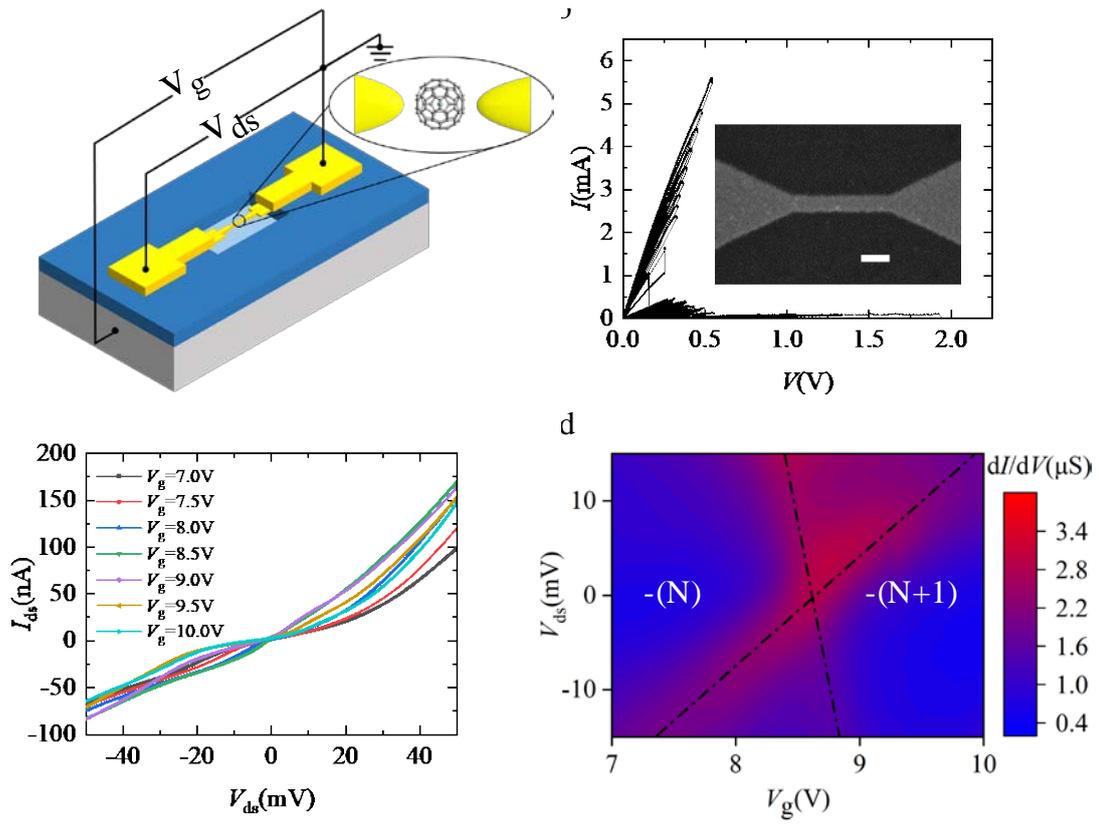

**Figure 1. Single electron transport of the Gd@C$_{82}$ SMD**

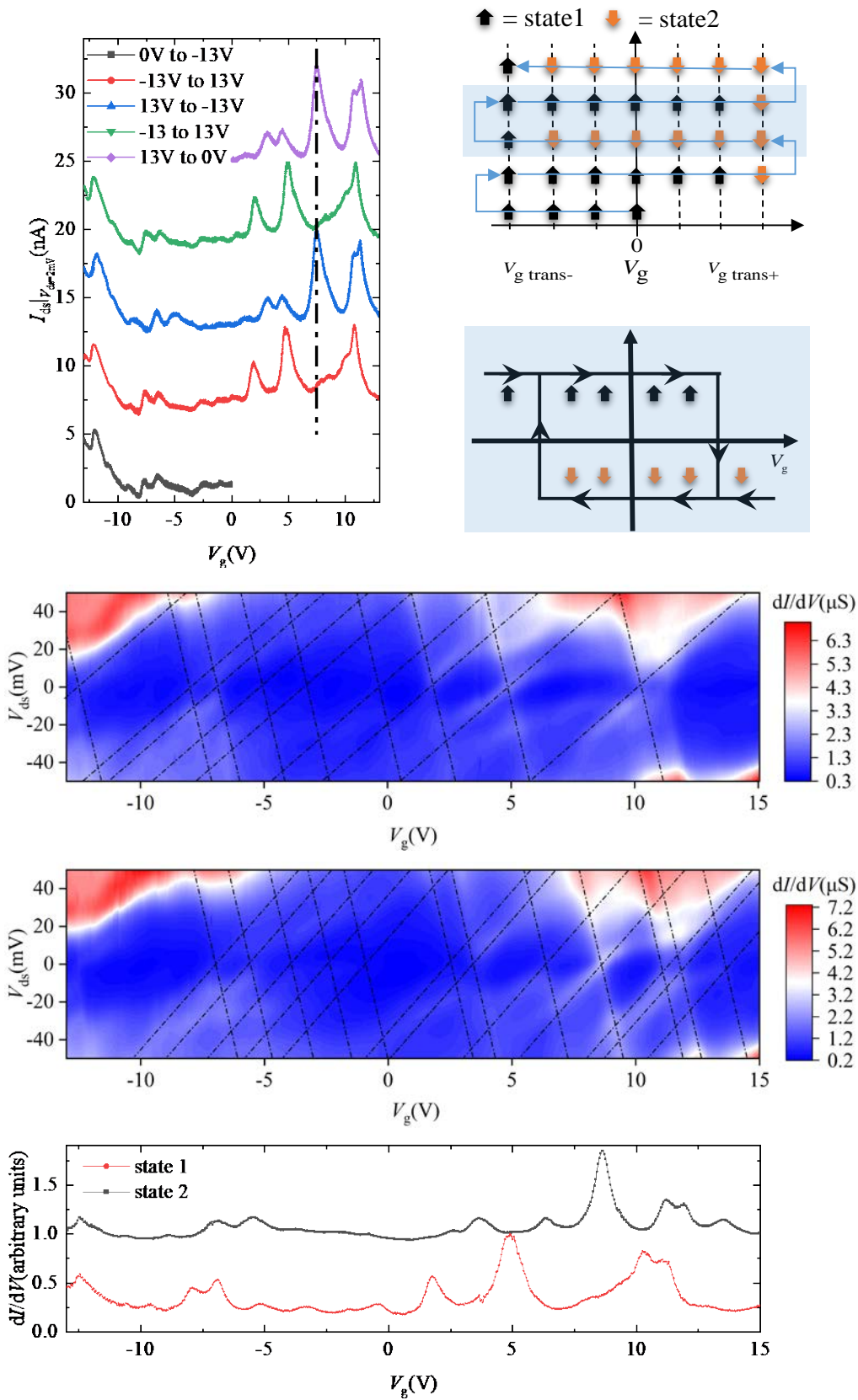

**Figure 2. Gate-controlled switching between the two molecular states showing a ferroelectricity-like hysteresis loop**

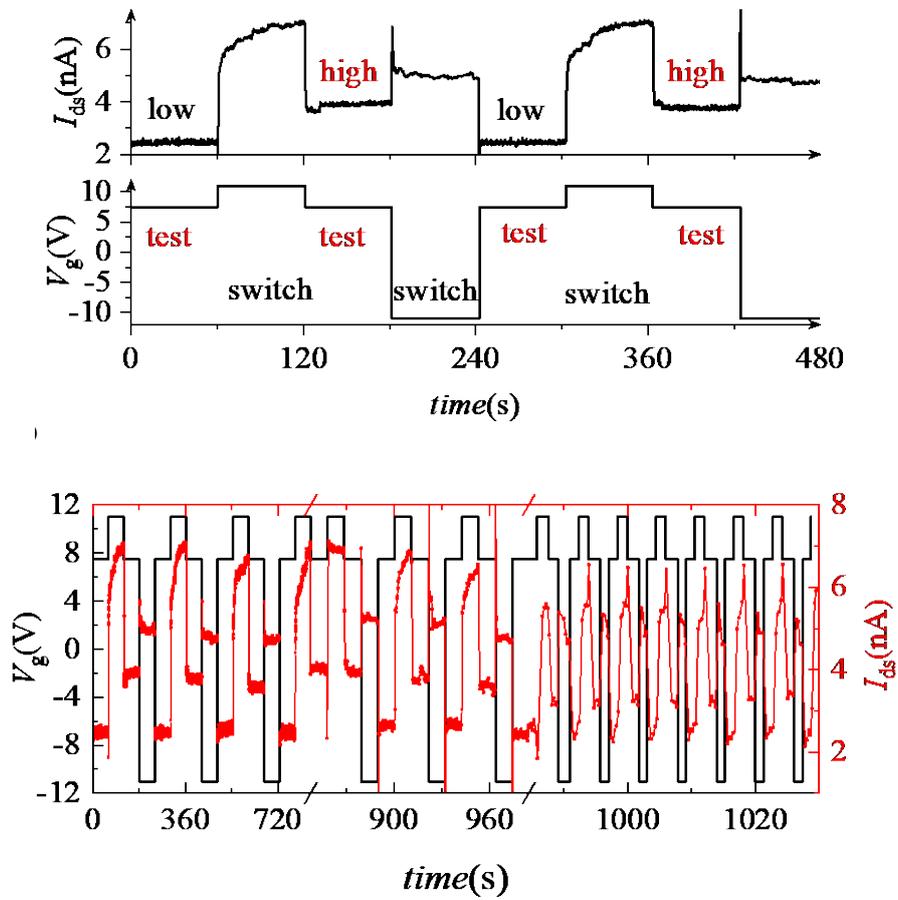

**Figure 3. Simulating a two-resistance-state operation based on the SMD switching**

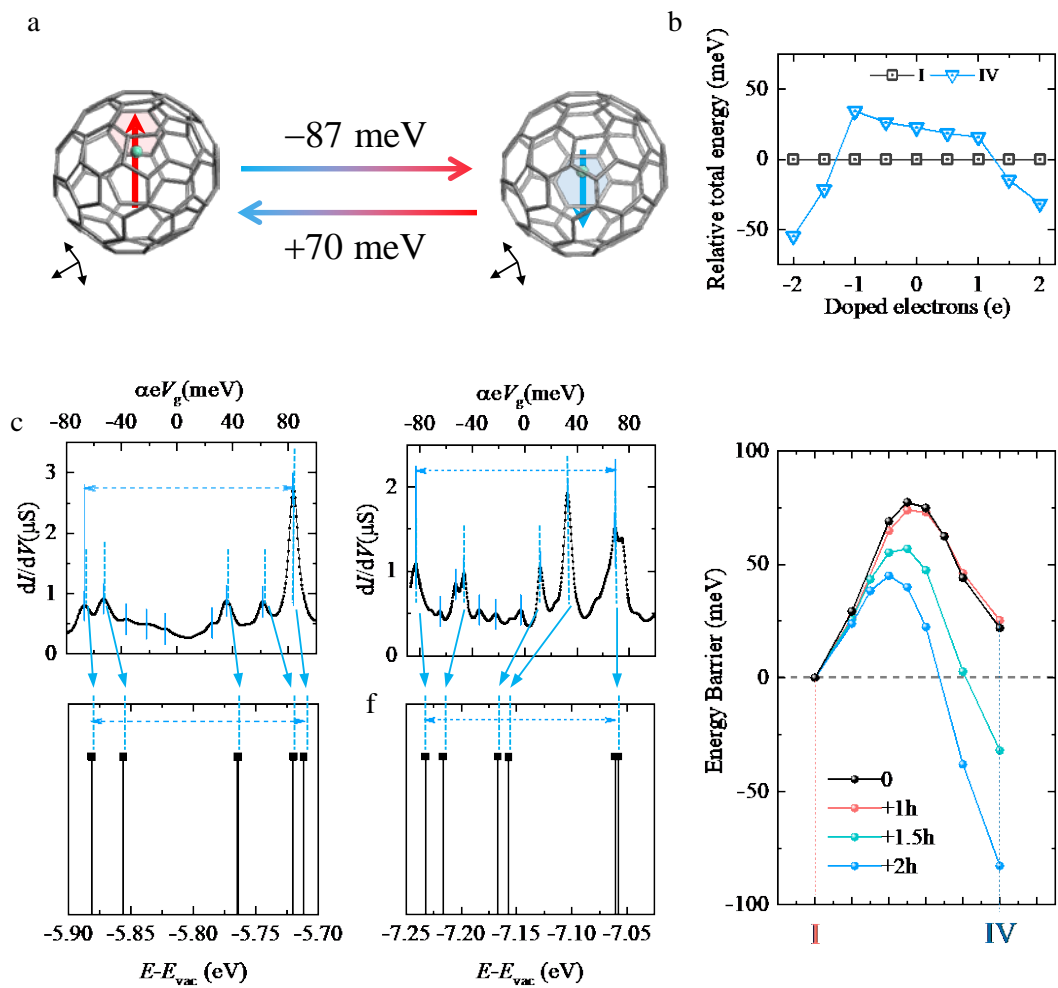

**Figure 4. Density functional theory calculations revealing the SME physics**

**SUPPLEMENTARY INFORMATION**

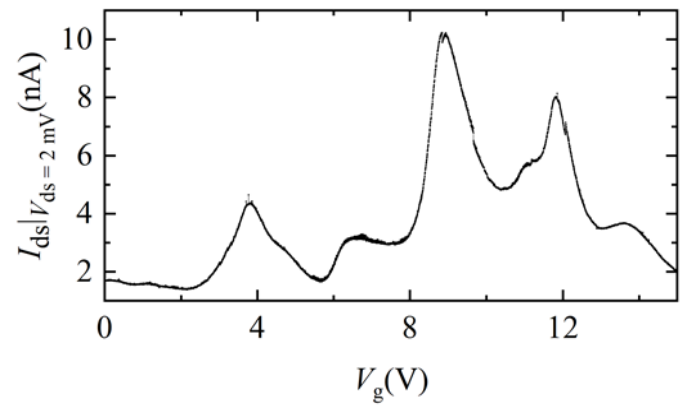

**Figure S1.** The source-drain current (fixing the bias voltage at 2 mV) measured as a function of the gate voltage, which has several peaks in the positive gate voltage regime, i.e. a series of charge degeneracy points.

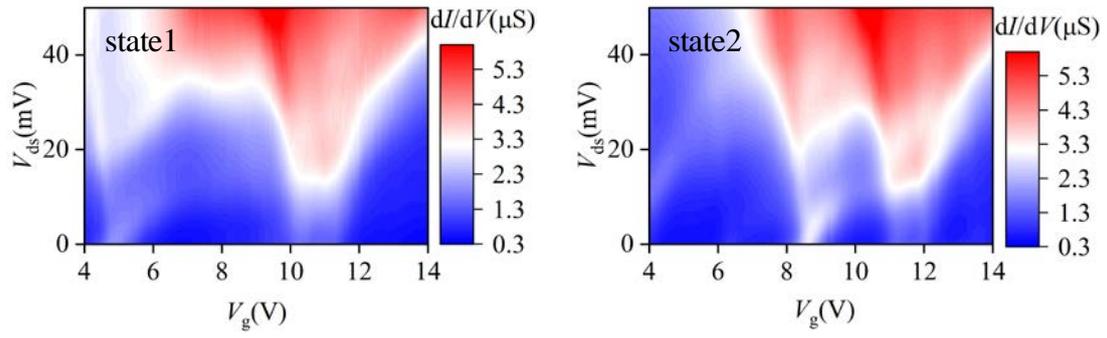

**Figure S2.** A color plot of the differential conductance d*I*/d*V* plotted against the bias voltage $V_{ds}$ and the gate voltage $V_g$. For clarify, selected bias voltage and gate voltage range come from Fig. 2(d) and 2(e), respectively.

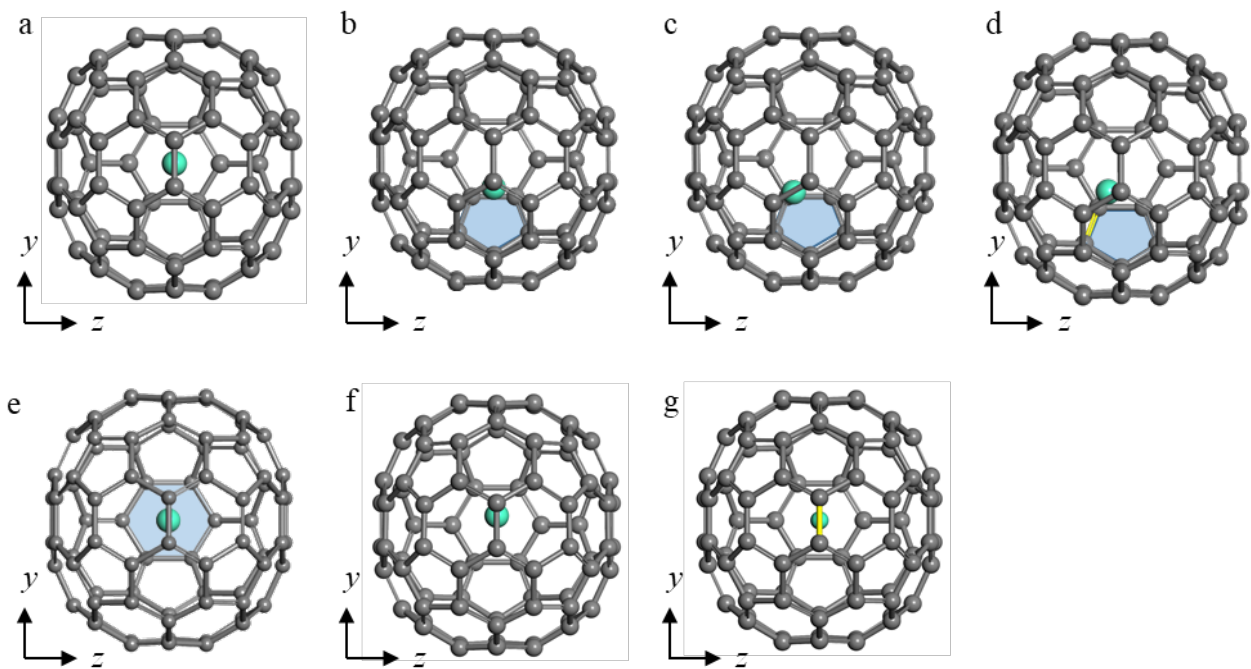

**Figure S3.** Seven adsorption sites of a single Gd atom adsorbed on the inner wall of a $C_{82}$ molecule were first considered in our calculation: (a) center of the molecule; (b-d) hollow (b), C-top (c) and bond-top (d) site of 5C-ring, respectively. (e-f) hollow (e), C-top (f) and bond-top (g) site of 6C-ring, respectively.

**Table S1**

Relative total energies and the heights of the adsorbed Gd atom at different sites. The Gd atom cannot stably hold its initial position on some of the adsorption sites during the relaxation process, and will transform to 6-hollow sites, which is marked with "→ 6C-Hol".

| Adsorption site | $\triangle E$ (meV) | $d_{C\text{-}Gd}$ (Å) |
|---|---|---|
| $C_{82}$-center | →6C-Hol | - |
| $C_5$-Hol | →6C-Gol | - |
| $C_5$-C-top | →6C-Hol | - |
| $C_5$-bond-top | →6C-Hol | - |
| **$C_6$-Hol-I** | **0.0** | 2.45 |
| $C_6$-C-top | →6C-Hol | - |
| $C_6$-bond-top | 1969.4 | 2.34 |

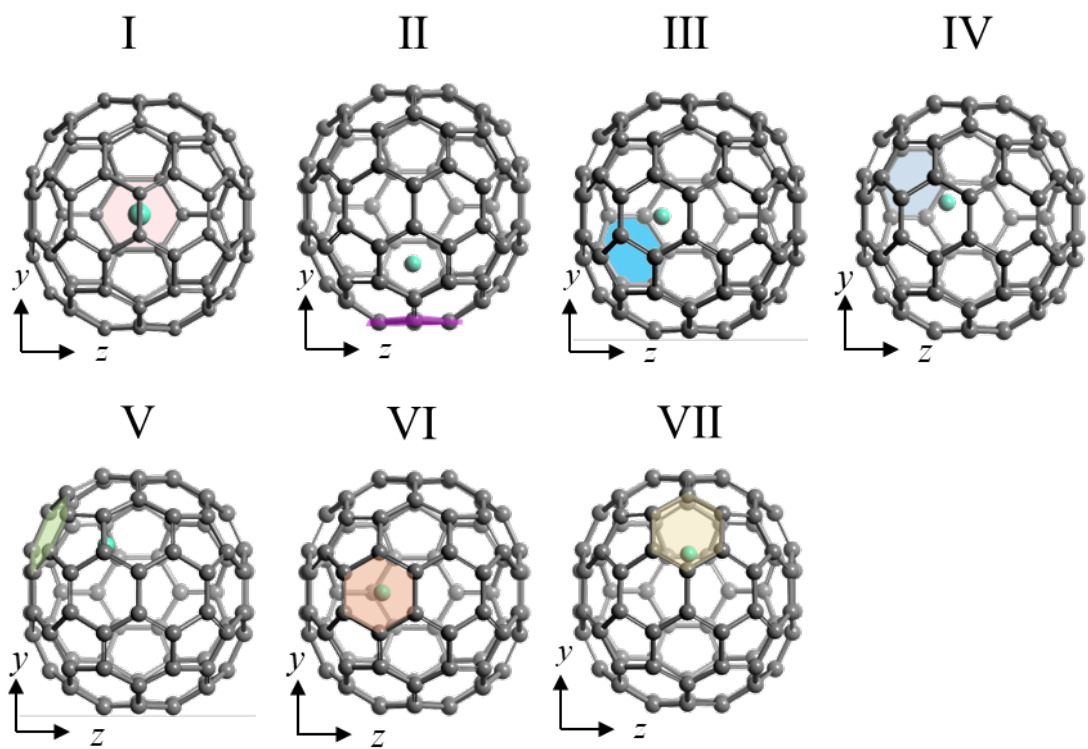

**Figure S4.** Seven configurations of inequivalent $C_6$ rings as Gd adsorption sites. The colorful filled hexagons represent the $C_6$ hexagonal faces with Gd atoms adsorbed.

**Table S2.** Relative total energies and the heights of the adsorbed Gd atom at the sites shown in Fig. S2

| Adsorption site | $\triangle E$ (meV) | $d_{C\text{-}Gd}$ (Å) |
|---|---|---|
| $C_6$-Hol-I | 0.0 | 2.45 |
| $C_6$-Hol-II | 474.9 | 2.38 |
| $C_6$-Hol-III | 65.8 | 2.33 |
| $C_6$-Hol-IV | 22.4 | 2.35 |
| $C_6$-Hol-V | 609.1 | 2.31 |
| $C_6$-Hol-VI | 1248.3 | 2.34 |
| $C_6$-Hol-VII | 1439.4 | 2.35 |

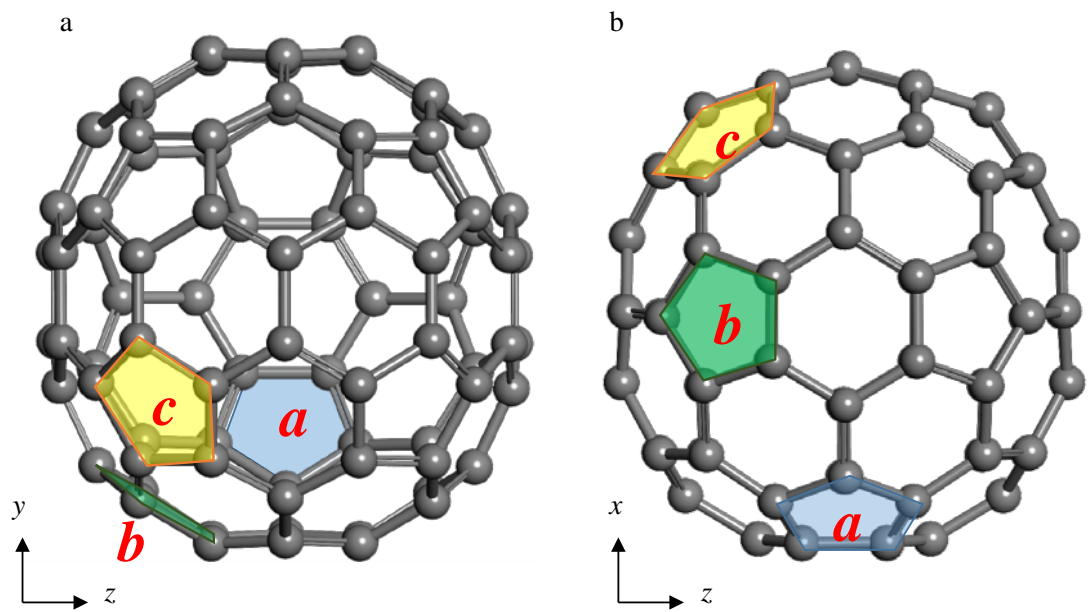

**Figure S5. Schematic of the three inequivalent C$_5$ rings adsorption sites considered in our calculation.** The light-blue, green and yellow filled pentagons represent the three C$_5$ faces a, b and c, respectively. For all the three C$_5$ rings, adsorption sites of hollow, C-top and bond-top were considered. Gd atom on all of these nine sites cannot stably hold its initial position and will transform to C$_6$-Hol or C$_6$-C-top sites.

**Table S3.** The calculated dipole moment of a single $C_{82}$ molecule, $C_6$-Hol configurations I, IV and 2.0 $h/C_{82}$ doped IV, respectively.

| Configuration | Dipole Moment $P$ (e*Å) | | | |
|---|---|---|---|---|
| | x | y | z | $|P|$ |
| $C_{82}$ | 0.38 | 0.00 | 0.00 | 0.38 |
| $C_6$-Hol-I | -0.50 | 0.00 | 0.00 | 0.50 |
| $C_6$-Hol-IV +2h | -0.36 | 0.19 | -0.32 | 0.52 |
| $C_6$-Hol-IV | -0.36 | 0.28 | -0.43 | 0.63 |